\documentclass[twocolumn,showpacs,preprintnumbers,amsmath,amssymb]{revtex4}

\usepackage{graphicx}
\usepackage{dcolumn}
\usepackage{bm}
\usepackage{amssymb}

\newcommand\muh{\mu^-  p \to   \nu_\mu n }

\newcommand\mum{\mu^- \to e^- \overline{\nu}_e \nu_\mu }
\newcommand\mup{\mu^+ \to e^+ \nu_e \overline{\nu}_\mu }

\def\address{\@ifstar{\address@star}%
  {\@ifnextchar[{\address@optarg}{\address@noptarg}}}

\begin{document}

\author{S.N.~Gninenko}

\affiliation{Institute for Nuclear Research, 117312 Moscow} 



\title{Muon capture rates from precision measurements of the muon disappearance }

\date{\today}

\begin{abstract}
In a typical  experiment the nuclear $\mu$-capture rate is determined from measurements of the  time constant of the muon decay exponential in a target. We propose a new approach for the evaluation of the $\mu$-capture rate,  which is based 
on  {\it direct} measurements of the $\mu^-$ disappearance rate in the target. 
Considering, as an example,  the reaction of $\mu$-capture on proton, we demonstrate that  measurements of the $\mu^-$ disappearance at different values of the hydrogen gas  pressure in the target could be used to avoid a pressure-dependent correction and determine the "vacuum" $\muh$ reaction  rate  with accuracy better than  $10^{-2}$ after  extrapolation to zero gas density. The proposed  method could be used to perform precision  measurements of the $\mu^-$ capture rate on different types of nuclei in one experiment.  
 \end{abstract}
\maketitle

\section{Introduction}

Precise measurements of muon capture rates  on nuclei 
are important to test concepts in nuclear and particle physics and to discover physics beyond the 
standard model of electroweak interactions,  see e.g. \cite{meas,kam}. 
 Among variety of possible reactions,
the muon capture on proton,  $\muh$, attracts significant attention \cite{kam1, aav}. 
The precise measurements of  the capture rate $\Lambda_S$ from the $\mu p$ singlet state in this  
process  is of fundamental interest  to determine the  pseudoscalar form factors in the 
axial nucleon current contributing to the week interaction between the muon and the proton.
The reaction $\muh$ has low rate  and its precision measurement  presents a challenge for the design and performance of the experiment.  
 The measurements are complicated by the fact that negative muons stopped in hydrogen 
could be captured not only from the $\mu p$ atomic state, but also from the $p p \mu $ molecules, where the 
muon capture rate differs significantly. 
The most precise results on the muon capture in hydrogen has been recently reported by the MuCap collaboration. For singlet rate they have obtained the value 
$\Lambda_S = (714.9 \pm 5.4_\text{stat.} \pm 5.4_\text{syst.})~ \rm{s^{-1}}$ \cite{mucap}, which corresponds to the branching fraction
\begin{equation}
Br(\muh)= \frac{\Gamma(\muh)}{\Gamma(\mu^- \to all)} \simeq 10^{-3}
\label{bratio}
\end{equation}
determined with the precision of $10^{-2}$, or the overall precision of $10^{-5}$ with respect to the muon decay rate.

 The MuCap detector was specially designed to significantly reduce the density-dependent formation of $p p \mu$ molecules by employing a low density hydrogen gas target. In this experiment, as well as in many others,  the nuclear $\mu$-capture rate was  determined from measurements of the  time constant of the muon decay exponential in the target.  
In this  paper  we discuss a novel approach allowing to improve the experimental precision of the muon capture rate on protons or other nuclei with a new type of measurements.  Instead of measuring the muon exponential  decay constant, we proposed to measure directly the 
 muon disappearance rate in the target. 
The rest of the paper is organized as follows. In Sec. II we describe the new approach, simulations of the signal, and the preliminary design of the experimental setup. The background sources and  the expected sensitivity  are discussed  in Sec. III and Sec. IV, respectively. Section V contains concluding remarks. 

\section{Precision measurements of the muon capture rate on protons}

\begin{figure*}
\includegraphics[width=0.9\textwidth]{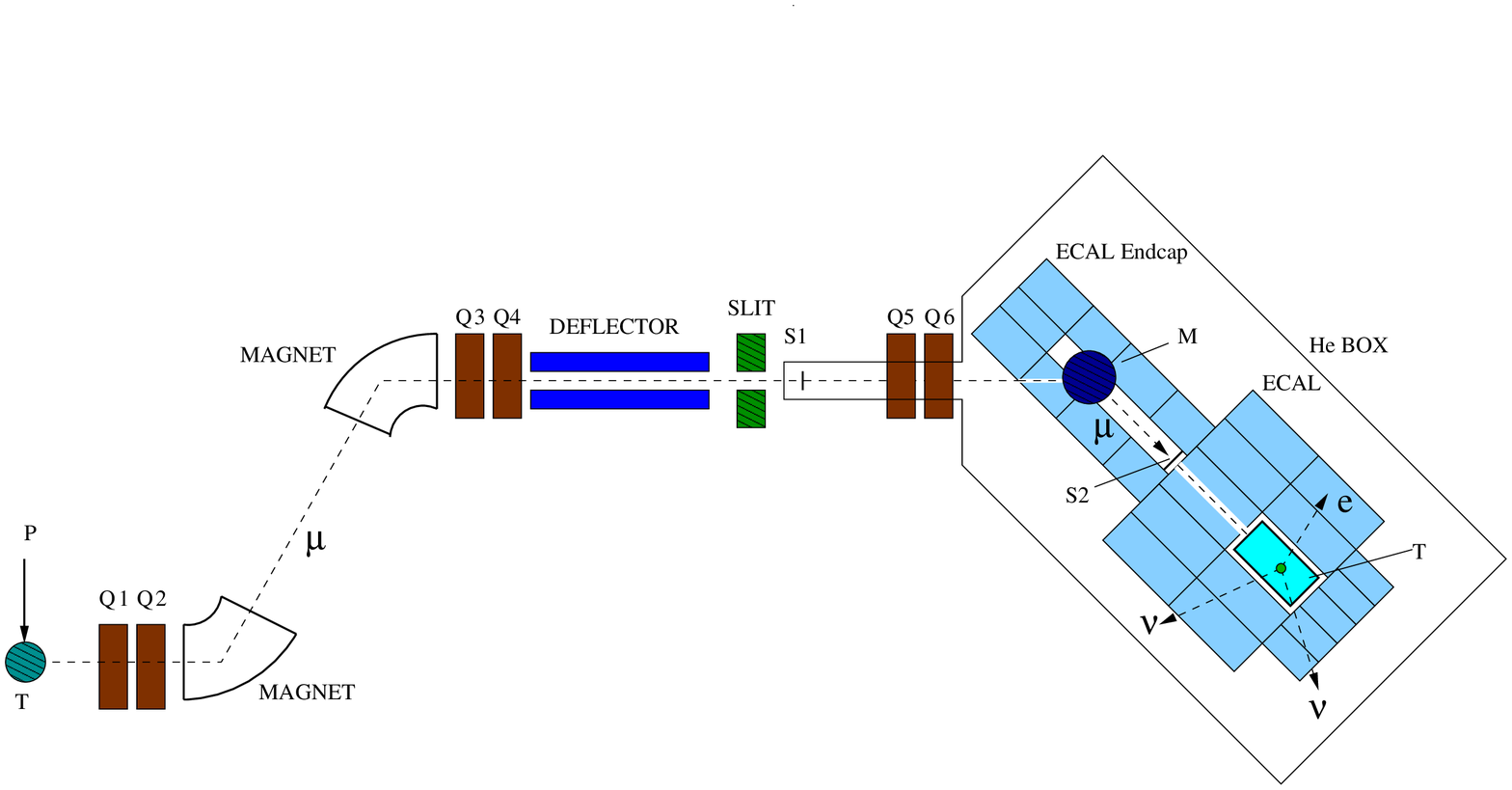}
\caption{\label{fig:setup} Schematic illustration of the experimental setup  for  precision measurements 
of the muon capture rate in protons.
 The $\mu^-$ low-energy beam passing through  the beam defining counters S$_{1,2}$
is focused by quadrupole magnets (Q5,Q6) into  a narrow aperture and  strikes the 
ultra-pure hydrogen gas  target  ($T$) used for the $\mu p$-atom formation. 
Shown are also the 4$\pi$ hermetic  BGO electromagnetic calorimeter (ECAL) and the magnet (M)  used to deflect the beam.
The counter S$_2$ and the upstream ECAL counters are also used as a veto against photons, decay electrons or backscattered muons that could escape the target through the entrance aperture. The deflector shown could be used 
to operate the setup in a "muon on request" mode.}
\label{setup}
\end{figure*}

To illustrate the method, consider an experiment on measurements of the  
muon capture rate on protons in the hydrogen gas target. 
The main components of the  experimental setup  are schematically illustrated in Fig. \ref{fig:setup},  see also \cite{sngmu, gkm}. The beam of surface $\mu^-$'s passing through 
the beam defining counters S$_{1,2}$ is focused  through a narrow aperture into a target  ($T$) used for the muonic hydrogen formation. The target is an Al vessel filled with high purity hydrogen. 
Shown are also the quadrupole magnets (Q) used for the beam focusing.  The energy of the beam is degraded by the counters material  to maximize the muon stopping rate in the target, where
 about 97\% of the muon captures occur in the $\mu p$ singlet state.
  The target is surrounded by a hermetic 4$\pi$ electromagnetic calorimeter 
(ECAL) to detect the energy deposited from the all muon processes in the target.  
 As shown in Fig. \ref{fig:setup}, before muons reach the entrance to the ECAL, they are bent in magnetic field.   
The purpose of utilizing the magnet is  to provide a transverse kick to negative muons in order 
to allow them to enter the target through the narrow aperture, and 
to detect photons, positrons, or muons that could escape the detection region through the entrance by a set of ECAL counters placed around the muon bend region, see e.g. \cite{gkm}. This additional detector is placed up stream of the entrance aperture. The counter S$_2$ is also used as a veto against  decay electrons or backscattered muons that could escape the ECAL through the entrance hole. The deflector could be used in order to operate the setup in a "muon on request mode" with the repetition rate of the order  100 kHz.
The  readout of the energy deposited in the ECAL  is triggered by a high efficiency tag signal of the muon 
appearance on the target,  defined as the  coincidence of 
signals from  the  counters S$_1$ and S$_2$ and   enhanced by using  the muon time-of-flight information.
To estimate the accuracy  of the experiment  a feasibility study  based on simplified GEANT4 \cite{geant}
 simulations combined with numerical calculations has been  performed.  
The  ECAL is an array of $\simeq 100$ bismuth germanate (BGO)   counters each of 
52 mm in diameter and  220 mm long, which was  previously used in the experiment searching for 
invisible decay of positronium \cite{bader}. The analysis takes into account active materials of the ECAL and passive materials from  
3 mm thick aluminum (Al)  vessel walls. 

\begin{figure}
\includegraphics[width=0.55\textwidth]{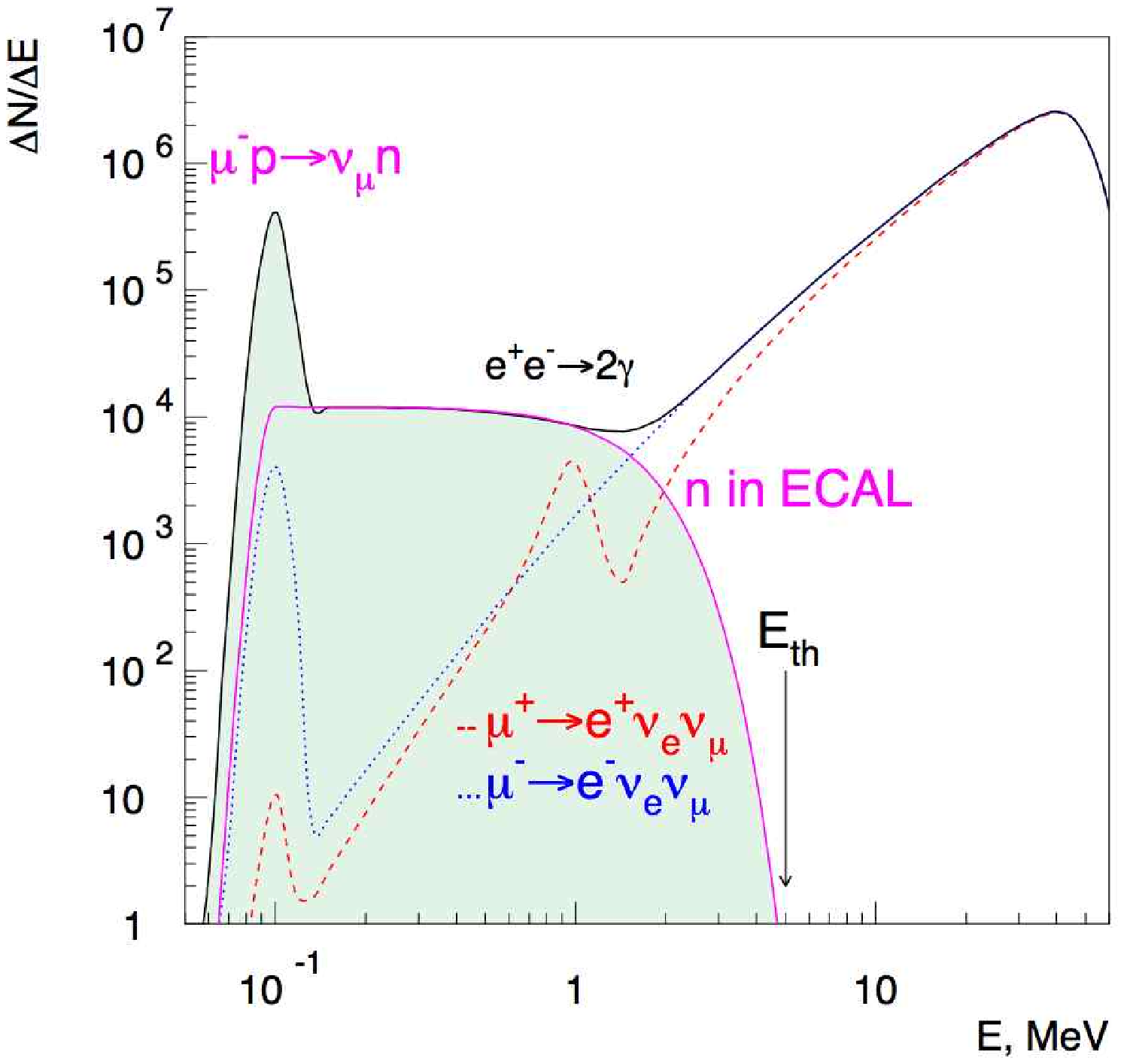}
\caption{ The expected distribution of energy deposition in the ECAL, after background subtraction, 
from  $10^{7}$ muons stopped in the hydrogen gas target, corresponding to the i) sum of  
decays $\mum$ and the reaction $\muh$ (solid curve). The  signal from the reaction $\muh$ (shaded area) corresponds to the rate $\Lambda_S \simeq 715$ s$^{-1}$; ii) the decay $\mup$ (dashed), and iii) pure Michel spectrum (dotted), shown for comparison.   
The peak around 1 MeV for the $\mup$ decays corresponds to 
energy deposition from the  $e^+e^- \to 2\gamma, 3\gamma$ annihilation of decay positrons stopped in the passive material.
The arrow shows the energy threshold for the process  $\muh$ detection.}
\label{spect}
\end{figure}

For  negative muons stopped in the target the expected 
distribution of the energy deposited in the  ECAL is given by a sum of four spectra:
\begin{equation}
n_\text{tot}(E)=\sum_i n_i f_i(E)  
\label{ntot}
\end{equation}
where $i= \mu, \nu,  n, b$, and  $n_i$ and $f_i(E)$ are the number of events and the normalized shapes of the energy distributions from the processes $\mum (\gamma)$, $\muh$ without ($i=\nu$) and with ($i= n $) recoil neutron energy deposition in the ECAL, and background,   respectively. 
The  background shape $f_b(E)$ including the ECAL counter's pedestals width is taken from simulations of  measurements  performed with a random trigger and assuming beam intensity of  $\simeq 10^4~\mu^-$'s/s, see, also Sec. III. 
The ECAL energy distribution obtained after background component subtraction  for 
 $\simeq 10^{7}$ $\mu^-$'s and $\mu^+$'s  stopped in the target is  shown in   Fig.  \ref{spect}.
For  the ordinary muon decay the experimental signature is the ECAL energy deposition
 from a single decay electron with energy $E_{e} = m_\mu - E_{\nu_e} - E_{\nu_\mu}$, where $  E_{\nu_e},  E_{\nu_\mu}$ are the electron and muon neutrino energy, respectively. The function $f_\mu$  is  calculated from the pure Michel spectrum convoluted with the  ECAL energy resolution \cite{bader}, and taking into account decay  electron interactions with the passive material.    

The ECAL energy  from the reaction $\muh$ originates from the recoil neutron interactions 
with nuclei of the ECAL matter, which are not or poorly detected. 
In this  case the neutron energy dissipates in a variety of mechanisms, resulting in the invisible final state described by the  function $f_\nu$. The neutrino, obviously, cannot be detected with any reasonable size calorimeter. 
The experimental signature of  the process $\muh$ with the apparent energy disappearance in the ECAL
 is an event with  the  sum of the ECAL crystal energies deposited by  the final-state particles equal zero. 
Zero energy is defined  in this case as an
energy deposition  below of a certain ECAL energy threshold,  $E < E_\text{th}$.   
Examples of such kind of measurements can be found in Ref.\cite{bader} describing the  experiment on a search for invisible decay of positronium, 
see also  \cite{opsmos}, or in a recent proposal on a search for the muonium annihilation 
$\mu^+ e^- \to \nu_e \overline{\nu}_\mu$ \cite{gkm}.     
 The function $f_{\nu}$ defined as "zero energy" in the ECAL, is parametrized by the Gaussian with  the
 $\simeq 50$ keV FWHM, in order  to reproduce the distribution of the ECAL counter pedestals \cite{bader}. 
The same feature, peak at zero energy ,  is also present in the Michel spectrum 
in the ECAL  corresponding to cases when decay electrons are not detected (see discussion below). 

A fraction of the $\muh$ reaction energy may be recovered when the recoil neutron is 
scattered or captured by other nuclei giving rise to a visible signal in the ECAL. 
 The ratio $n_\nu / n_n$ depends on the ECAL
crystal type. For BGO crystals it is  of the order $\simeq 10 \%$ \cite{nbgo}.
 The function $f_n$ is parametrized by using the BGO crystal prompt response to 
 4 - 6 MeV neutrons from Ref.\cite{nbgo} with a proper rescaling. The uncertainties of the shape of the 
 $f_n$-distribution are
 not very important (actually, the function $f_n$ can be measured in the proposed  experiment).
 The most  important  fact is  that the neutron energy spectrum  vanishes at  $\simeq$ 5 MeV, i.e. it is 
 below the energy threshold of  $E_\text{th}\simeq 5$ MeV.
 So, one can consider the energy region $E \lesssim 5$ MeV as the signal region for the 
 reaction $\muh$. The higher energy part of the spectrum, above $\gtrsim$ 5 MeV, is assumed to be not affected  by the  energy deposition from the   recoil neutron in the reaction $\muh$.
 It is  calculated from the pure Michel spectrum convoluted with the  ECAL energy resolution.

As shown in Fig. \ref{spect}, the $f_{\nu,n}$ and $f_\mu$ spectra are not completely separated.
In order to extract the number of signal events 
 and determine the rate $\Gamma_{\mu p}$ of the reaction $\muh$, one has to  measure accurately the  corresponding number $n_\mu$ of background events  from the decay $\mum$ in the region $E \lesssim 5$ MeV:
\begin{eqnarray}
&&n_s =n_{\nu}  + n_{ n}\nonumber \\ 
&&= n_\text{tot}(E<E_\text{th}) - n_\mu(E<E_\text{th})  
\label{nsig}
\end{eqnarray}
 This inevitable  background came from the muon decay into a final state 
 electron which after energy loss in the vessel material either completely stop in it or  enters  the ECAL with the  
 kinetic energy $E_\text{kin} \lesssim$ 5 MeV. In both cases  the event results in a fake $\muh$ signal.
 The partial muon decay rate $\Delta \Gamma_\mu$ 
into an electron with  $E_{e^-} < E_\text{th}$  as a function of $E_\text{th}$ is given by 
\begin{equation}
\Delta \Gamma_\mu (E<E_\text{th})\simeq 16 E^3_\text{th}/m^3_\mu, 
\label{partial}
\end{equation}
where $m_\mu$ is the muon mass.
To reduce the number of such background events one has to use  as low as possible threshold  $E_\text{th}$.
Taking into account that energy of positrons   that stop in the Al vessel  is  typically   $E_\text{kin} \lesssim  2$ MeV, the  fraction of stopped $e^-$'s is estimated to be $\lesssim 10^{-4}$.   The target vessel should be optimized in size  by keeping the amount of passive material  as small as possible.

The contribution of low energy electrons  from the muon decay in the signal region,  can be obtained by using spectrum of the energy 
deposition in the ECAL from the decays $\mup$ of positive muons. The decay rate of positive muons that stopped in the target is not affected by the capture reaction. Hence, assuming the initial identity of the $e^-$ and $e^+$ spectra, the fraction of low energy positron in the signal region can be precisely measured and used for the evaluation of the corresponding fraction of 
electron  events from the decay $\mum$.   
 
  The calculated distribution of the energy deposition in the ECAL from decay positrons is shown in Fig. \ref{spect}. The spectrum is 
 a superposition of two spectra: one from positrons that stopped and annihilated in the target, another one corresponds to positrons that struke the ECAL. The letter is shifted to  higher energies by the amount of additional energy $\simeq 1$ MeV, from the positron  annihilation in the BGO crystals. 
Positrons that stopped in the vessel,  differently from the electrons, produce visible signal due to their annihilation into 2 or 3 $\gamma$'s at a lifetime scale of the order of a few ns. 
One can  see a peak around 1 MeV in Fig. \ref{spect} from these events, whose fraction depends on the Al vessel thickness.
As the attenuation length for 511 keV annihilation $\gamma$'s in Al is much smaller then that of $\simeq$ MeV electrons ($\simeq$ 3 cm for $\gamma$'s, and  $\simeq$ a few mm for e$^-$'s with energy of $\simeq$ 1 MeV), most of $\gamma$'s are  detected. Simulations show that the main contribution to the $\gamma$-detection inefficiency comes from the total (due to photo-absorption) or fractional 
(due to Compton effect) photon energy loss in the material of the vessel. 
Therefore, it is preferable to have the vessel  made of a low-Z material to minimize the e$^\pm$'s energy loss and the cross-section of the photo-absorption,  which is $\sim Z^5$. For the vessel thickness of 3 mm, however, only a  fraction of $\ll$ 1\% of annihilation photons would be completely absorbed. 
A negligibly small  peak at zero energy in Fig. \ref{spect} represents such losses of annihilation energy. 
Thus, for the most positrons stopped in the vessel, the energy deposited in the ECAL 
would be $ \simeq 1$ MeV, making these  events visible, as illustrated in  Fig. \ref{spect}. 

The fraction of $e^\pm$'s events from muon decays with energy below the  threshold $E_\text{th}\lesssim 5$ MeV is $\simeq 10^{-3}$. In order to be competitive with the reported precision
  of Eq.(\ref{bratio}), one has to determine the contribution from   $n_{\mu^-}(E< E_\text{th})$ to the signal region  with accuracy at least $\simeq 10^{-2}$ or better. As discussed previously, this  number $n_\mu (E<E_\text{th})$ of events from $\mu^-$ decays can be  determined by using the proper normalized spectrum of the energy 
 deposition in the ECAL from positive muon decays. For a $E_\text{th}$ value far from the annihilation peak, see Fig. \ref{spect}, the corresponding numbers of $e^-$ and $e^+$ events are assumed to be connected by the relation 
\begin{equation}
n_{\mu^-}(E< E_\text{th}) = n_{\mu^+}(E<E_\text{th}+1.022~{\rm MeV}).
\label{events}
\end{equation}  
Below, we discuss the accuracy up to which  Eq.(\ref{events}) is valid, and show that it is suitable for the purpose of the experiment. 
The statistical error at the level $\simeq 10^{-3}$ in determining of $n_{\mu^-}(E< E_\text{th})$ from Eq.(\ref{events}), could be easily achieved during one day of running the experiment at a  moderate intensity $I_\mu \simeq 10^4~\mu$/s. 
Several sources of systematic uncertainties, such as possible difference between the ECAL responses to the energy deposited by electrons, positrons and photons,  backscattering from the BGO crystals, etc.. have been considered.  
The dominant source of systematic uncertainties is found to be due to a small  difference between the electron and positrons energy loss in matter. For the  energy region $E \gtrsim 5$ MeV, the ECAL energy spectra of  $e^+$'s and $e^-$'s  are related by   
\begin{equation}
f_{\mu^-}(E) = f_{\mu^+}(E+\delta(E)+1.022~{\rm MeV})
\label{corr1}
\end{equation}
where the correction $\delta(E)$  appears  due to the difference between the $e^+$'s and $e^-$'s energy loss 
rate in the vessel. Comparison of positron and electron energy loss in Al shows that 
for the energy range $5\lesssim E \lesssim 50$ MeV the collisional energy loss ratio is in the range \cite{epdiff}:
\begin{equation} 
0.971 < [(dE/dx)_{e^+}/(dE/dx)_{e^-}]_\text{coll} < 0.973, 
\end{equation}
 while for the total energy loss it is 
\begin{equation} 
0.977 < [(dE/dx)_{e^+}/(dE/dx)_{e^-}]_\text{tot} < 0.986.
\label{corr2} 
\end{equation}
Simplified simulations of $\simeq 10^7$ events were
performed to  obtained the $e^-$ and $e^+$ spectra. The $e^+$ ECAL energy distribution was corrected for the
energy loss difference by using Eqs.(\ref{corr1},\ref{corr2}) and normalized to the $e^-$ spectrum in the  energy region $E \gtrsim 5$ MeV. For $10^7$ $\mu^\pm$ decays, the ratio of the number of background electron events in the signal region $E\lesssim 5$ MeV, determined from the $\mu^+$ decay spectrum,  to the  number of true background electron events were found to be  $n_{\mu^+}(E< E_\text{th})/n_{\mu^-}(E< E_\text{th}) = 0.985\pm0.012$. The error in this estimate  is defined by the statistical one combined with the uncertainty  of corrections in Eq.(\ref{corr2}), taken to be $\pm 0.001$. 
If one extrapolates this result to the total number of  $ \simeq 10^9$ events 
accumulated during one day of running,  then, the number of background electron events in the $\muh$ signal region can be predicted  with a  precision $\lesssim 10^{-2}$.  Although such extrapolation may  be imperfect and the question -
have all sources of systematic errors been properly estimated?- still remains,  we do not see a potential source of systematic uncertainty which prevents us from doing this. Indeed, even if we neglect the $e^+$-$e^-$ difference and take $\delta=0$ in 
Eq.(\ref{corr1}), the $e^-$ and $e^+$ spectra are systematically shifted with respect to each other by $\simeq 0.03$ MeV in the energy region $E\gtrsim 5$ MeV.  Taking into account Eq.(\ref{partial}), results in $\simeq 2\%$ systematic error in the prediction of $e^-$ background events due to this energy shift, which is  comparable with the required precision.  
   Therefore, this procedure of   background estimate from   low energy electrons allows  potentially to reach accuracy  in the muon capture rate  as small as $\lesssim 10^{-2}$.

In order to cross-check simulations of  $\mu^+,~ \mu^-$  spectra features, one could also compare  the number of electrons stopped in the passive material with those predicted from  the annihilation peak in the positron spectrum. For this purpose, the  region around the peak is described by a function  $n_1 \cdot f_1(E_{e^+}) + n_2\cdot f_2(E_{e^+})$, which  is a  sum of a distribution corresponding to  the 1 MeV peak from the positron annihilation in the ECAL and  a  polynomial background.  The fit results in prediction of $4579\pm 71$ background $e^-$ events contributing to the $\muh$ signal region, which is found to be in a good agreement with the true number of 4503 $e^-$'s stopped in the target.  

 \section{Background}

The background processes for the $\muh$ reaction can be classified as being due to 
 physical, detector-, and  beam-related backgrounds. To perform the full detector simulation and 
to investigate  these backgrounds down to the level  $\simeq 10^{-6}$ would require the generation of a large number of muon decays resulting in a large amount of computer time. Consequently, background processes are estimated  with a  smaller statistics combined  with numerical calculations.
 
 The following sources of physical- and detector-related backgrounds are considered. The first one is related to the incomplete ECAL hermiticity. Our  study   identified this background as due to energetic decay products, $e^\pm$ and $\gamma$'s,  escaping the detection region  
 though the entrance aperture. This process increases the 
 disappearance rate of the muon and hence must be addressed. If a muon decays in flight in the target into a fast 
electron (positron)  with momentum pointing  to the ECAL entrance hole it also contribute to the signal region.
The feasibility study of the experiment on the muonium decay into neutrino, shows that this  contribution from the incomplete ECAL hermiticity is well under the level $\lesssim 10^{-6}$;    
for more detail  discussions, see  Ref.\cite{gkm}. For the Al vessel thickness of 3 mm, background due to the annihilation photons absorption in the vessel material is found to be small. 
 
 The  beam-related backgrounds produce the fake muon tag and   can be classified  as being
due to a  beam particle misidentified as a muon, or several 
 beam particles  which produce fake muon tag due to 
accidental coincidence of signals from S$_{1,2}$. Identification of the incoming
particle as a muon based on the  requirements of the delayed by the muon time-of-flight 
coincidence between the beam counter signals 
suppresses the single-beam background down to the negligible level.
The performed analysis shows that the beam-related background is dominated by the pileup events.
The pileup energy, which corresponds to extra  energy deposited in the ECAL counters  by an additional undetected and uncorrelated particle, increases values of the counters pedestals and  shifted them into 
higher energy region. The amount of additional energy in each ECAL  counter can be accurately  measured with the random trigger provided by an external clock of low frequency \cite{bader} and then subtracted from the energy  spectrum, as discussed previously.  
\begin{figure}
\includegraphics[width=0.5\textwidth]{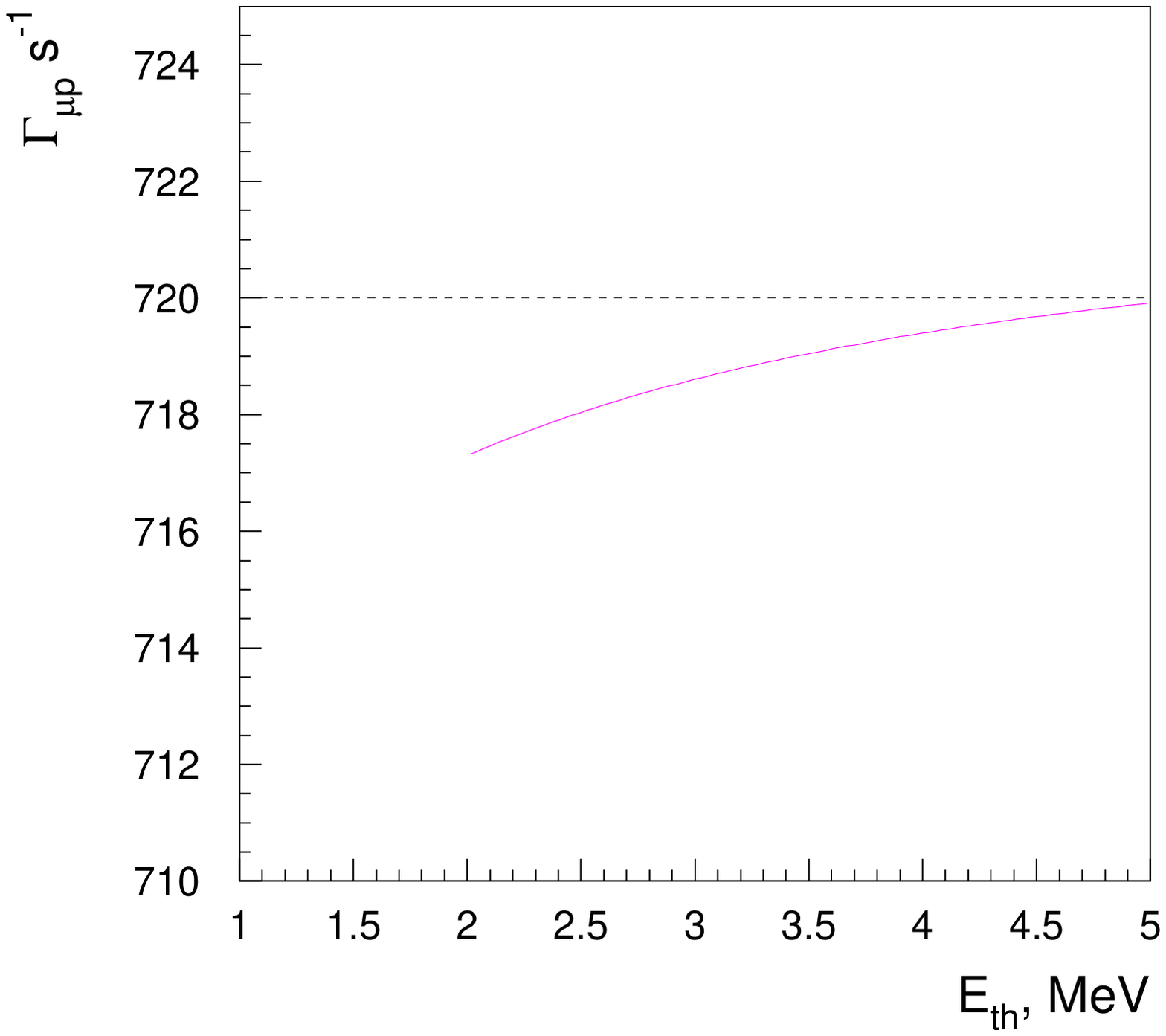}
\caption{ The expected dependence of the $\Gamma_{\mu p}$ rate as a function of energy threshold for the given 
pressure in the hydrogen gas target. The vacuum value of $\Gamma_{\mu p}$ rate
can be obtained from the extrapolation of its values measured at different gas pressures to zero gas density.}
\label{lambda}
\end{figure}

\section{Expected results}

The number of observed  $\muh$ signal events  for a given hydrogen gas pressure $p$ in the target is given by 
\begin{equation}
n_s = \frac{\Gamma_{\mu p}(p)}{\Gamma_\mu+\Gamma_{\mu p}(p)} n_\text{tot}
\label{rate}
\end{equation} 
where $n_\text{tot}=I_\mu t$, and  
$t$ is the  running time of the experiment. The $\Gamma_\mu = 1/\tau_\mu$  is taken from 
the high precision measurements results of the positive 
muon lifetime $\tau_\mu$ recently reported by the MuLan collaboration \cite{mulan1,mulan2,mulan3}:
\begin{equation}
\tau_\mu = 2.1969812\pm 0.0000038 ~\mu{\rm s}
\label{tauq}
\end{equation}
and assuming that the lifetimes of $\mu^+$ and $\mu^-$ are equal to each other.
Note, that the muon lifetime found in these measurements is obtained with precision $\simeq 1$ ppm. 
The muon disappearance rate in the detector is given by \cite{mucap} 
\begin{equation}
\Gamma_{\mu p}(p) = \Delta\lambda_{\mu p} + \Lambda_S +  \Delta\Lambda_{ p p \mu} (\rm p)
\label{ratep}
\end{equation}where $\Delta\lambda_{\mu p} $ is a calculable correction due to the $\mu p$ bound-state effect \cite{c1,c2}
and  $\Delta \Lambda_{ p p \mu}(\rm p)$ is pressure depending correction, which takes into account a fraction of muons captured from molecular states. In the MuCap experiment the latter corresponds to about 3\% of muons. 
 The correction $\Delta \Lambda_{ p p \mu}(\rm p)$ depends on the muon capture rate from orth- or para-$pp\mu$ state, 
  and on the ortho-to-para  transition rate.  It could be derived  from fits to simulated data \cite{mucap}. This evaluation, however, is known with a limited accuracy. 
 
 A more
appropriate way  to obtain the  $\Gamma_{\mu p}(0)$ rate  would be the  extrapolation of $\Gamma_{\mu p}(p)$  
obtained from measurements at several different values of the hydrogen gas pressure in the target  to zero gas density.  In Fig. \ref{lambda} the dependence of the  $\Gamma_{\mu p}$ value extracted from simulated spectra of Fig.\ref{spect} by using  Eq.(\ref{rate}) 
as a function of the ECAL energy threshold for a given hydrogen gas pressure is shown for illustration. 
Assuming the  muon intensity of $I_\mu\simeq 10^{4}~ \mu^- /$s  and using
 Eq.(\ref{bratio}),  we anticipate   $\simeq   10^{6}$ $\muh$ signal events 
per one day of running the experiment. The estimate shows, that  the "vacuum" $\muh$ capture  rate  can be  obtained 
by a 0.2\% extrapolation of six measured values at the gas pressure from 5 to 10-bar to zero gas density. Here, we use linear 
extrapolation, assuming that for each pressure value the  $\Gamma_{\mu p}$ rate is measured with precision $\pm 0.5$\%, and the 
formation rate of $pp\mu$ states is proportional to the gas density.
Clearly, the statistical accuracy can be easily improved with more accumulated data.
The extrapolation will remove pressure-dependent systematic effects related to the $\mu p $ reaction from molecular states in the target. Hence, this experiment is virtually a  direct measurement of the "vacuum" muon capture rate on protons. However, it is a  subject
to very different systematic effects than in the MaCap experiment, as discussed above.
In Table I contributions from the previously discussed processes to the uncertainty of the 
 $\muh$ capture rate determination  are summarized. 
The dominant  sources are expected to be due to the small difference between the electron and positron  energy loss in the vessel material, uncertainties of the extrapolation procedures and of   
the recoil neutron energy distribution in the ECAL.
\begin{table}[htb!]
\caption{\label{tab:table1} Expected contributions to the accuracy of the $\muh$ reaction rate  
measurement  from different background sources ( see text for details). }
\begin{ruledtabular}
\begin{tabular}{lr}
Source of uncertainty & Expected level\\
\hline
fake $\mu^+$, $\mu^-$ tag & $ \lesssim 10^{-6}$\\
$e^+$, $e^-$ difference & $ \lesssim  10^{-2}$\\
ECAL hermiticity  & $\simeq 10^{-5}$\\
recoil neutron ECAL spectrum & $\simeq 10^{-3}$\\
extrapolation vs $E_{\text{th}}$ & $\simeq 10^{-3}$\\
extrapolation vs gas pressure & $\simeq 10^{-3}$\\
\hline 
Total   & $\lesssim  10^{-2}$\\
\end{tabular}
\end{ruledtabular}
\end{table} 
Now,  let us discuss several additional limitation factors.
The first one is related to the relatively long muon lifetime. In order 
 to get  the measurement precision of the branching fraction $Br(\muh)$ of the order  $\simeq 10^{-6}$, the ECAL gate duration $\tau_g$, and hence the dead-time per trigger, has to be 
\begin{equation} 
\tau_g \gtrsim - \tau_\mu \times ln(Br(\muh)) \simeq 30 ~\mu s
\label{dur}
\end{equation}
 in order to avoid background from the muon decays outside the gate. 
In the ETH-INR positronium experiment, the ECAL gate $\tau_{Ps}$ 
was about $\simeq 2~\mu s$ for orthopositronium lifetime in the target of 132 ns. This 
resulted in distribution of the sum of pedestals of all ($\simeq 100$) 
ECAL counters corresponded to the threshold of 80  keV used to define the signal range for the 
$o-Ps \to invisible$ decay. In the proposed  experiment the
 longer gate will lead to  an increase of the pile-up and pick-up electronic 
noise and hence to the overall
broadening of the $f_\nu$ signal,  approximately by a factor  
$\sqrt{\tau_g/\tau_{Ps}}\simeq 4$
and, hence to an increase 
of the energy threshold roughly up to $E_\text{th}\simeq 300$ keV  \cite{bader}. 
For this threshold the probability of the  energy loss from the positron annihilation is 
about $P_{2\gamma} \simeq 10^{-6}$ \cite{bader}, which is still comparable with the  expected overall sensitivity 
of the experiment.

Another limitation factor is related to the dead time of  Eq.(\ref{dur})
and, hence to  the maximally allowed muon counting rate, which  according to  
Eq.(\ref{dur}) has to be  $ 1/ \tau_g \simeq  10^{4} \mu / s$ to avoid significant pile-up effect.    

\section{Summary}
In summary, 
in a typical  experiment the nuclear $\mu$-capture rate is determined from measurements of the  time constant of the muon decay exponential in a target. Here, we propose a new technique for the evaluation of the $\mu$-capture rate,  which is based 
on  {\it direct} measurements of the $\mu^-$ disappearance rate in the target. As an example, we consider 
 the reaction of $\mu$-capture on proton, and show that measurements of the $\mu^-$ disappearance at different values of the hydrogen gas  pressure in the target allow to avoid a pressure-dependent correction and determine the vacuum $\muh$ reaction  rate  with accuracy better than  $10^{-2}$ after  extrapolation to zero gas density. This experiment is a subject to very different systematic effects than, e.g.,  in the recent MuCap experiment \cite{mucap}.  The  quoted sensitivity could be obtained with a setup
optimized for several its properties. Namely, i) the energy
resolution, material composition and dimensions of the
target vessel, ii) the efficiency of the beam (veto) counters, and iii) the
pile-up effect and zero-energy threshold in the ECAL , are of importance. The question -how reliable are simulations of the 
difference between electron and positron energy loss and  response of the ECAL - might require further study.    
The described technique could be  used for precision measurement of the muon 
capture rate in cold deuterium gas, which is  the main goal of the MuSun experiment  \cite{musan}.
Since these measurements  have important astrophysical implications, it would be interesting to perform independent experiment with different methods, by using, for example, a setup, which has been  recently  proposed to search for the muonium annihilation 
into two neutrino at the Paul Scherrer Institute \cite{gkm, loi}.  
A similar detector without high pressure gas  requirements is simpler and could be used for precise measurements with proposed method 
 of the nuclear muon capture rate  with different  targets in one experiment.

Finally note, that the reported analysis  gives an illustrative correct order of magnitude for the precision  of the 
proposed method of muon capture rate measurements  and may be strengthened  by more accurate and  detailed Monte Carlo 
simulations of the concrete  experimental setup, which are beyond the scope of this work.

\begin{acknowledgments}
I am grateful to D. Sillou for usefull comments and his help in calculations. The members of the PSI Advisory Committee are 
gratefully acknowledged for their valuable comments on type of the measurements described in this work.
\end{acknowledgments}

\end{document}